\documentclass[12pt]{JHEP3}

\usepackage{amsmath}
\usepackage{amsfonts}
\usepackage{mathrsfs}
\usepackage{cite}

\newcommand{\beqs}{\begin{equation*}}
\newcommand{\beq}{\begin{equation}}

\newcommand{\eeqs}{\end{equation*}}
\newcommand{\eeq}{\end{equation}}

\newcommand{\beqas}{\begin{eqnarray*}}
\newcommand{\beqa}{\begin{eqnarray}}

\newcommand{\eeqas}{\end{eqnarray*}}
\newcommand{\eeqa}{\end{eqnarray}}

\newcommand{\eq}[2]{\begin{equation} #1 \label{#2} \end{equation}}

\newcommand{\eps}{\varepsilon}
\newcommand{\al}{\alpha}
\newcommand{\be}{\beta}
\newcommand{\ga}{\gamma}
\newcommand{\de}{\delta}

\newcommand{\la}{\lambda}
\newcommand{\si}{\sigma}

\newcommand{\Ga}{\Gamma}

\newcommand{\blist}{\begin{itemize}}

\newcommand{\elist}{\end{itemize}}

\providecommand{\href}[2]{#2}

\DeclareFontFamily{OT1}{rsfs}{}
\DeclareFontShape{OT1}{rsfs}{m}{n}{ <-7> rsfs5 <7-10> rsfs7 <10->rsfs10}{} 
\DeclareMathAlphabet{\mycal}{OT1}{rsfs}{m}{n}

\newcommand\MM{\mathcal{M}}
\newcommand\dM{\partial \MM}

\title{Instability in cosmological topologically massive gravity at the chiral point}

 \author{Daniel Grumiller\\
           Center for Theoretical Physics, 
           Massachusetts Institute of Technology,\\
           77 Massachusetts Ave.,
           Cambridge, MA  02139\\
           Email: \email{grumil@lns.mit.edu}}

 \author{Niklas Johansson\\
          Institutionen f\"{o}r Fysik och Astronomi, Uppsala Universitet \\
          Box 803, S-751 08 Uppsala, Sweden\\
          Email: \email{Niklas.Johansson@fysast.uu.se}}

\abstract{
We consider cosmological topologically massive gravity at the chiral point with positive sign of the Einstein--Hilbert term. We demonstrate the presence of a negative energy bulk mode that grows linearly in time.
Unless there are physical reasons to discard this mode, this theory is unstable. To address this issue we prove that the mode is not pure gauge and that its negative energy is time-independent and finite. The isometry generators $L_0$ and $\bar L_0$ have non-unitary matrix representations like in logarithmic CFT. While the new mode obeys boundary conditions that are slightly weaker than the ones by Brown and Henneaux, its fall-off behavior is compatible with spacetime being asymptotically AdS$_3$.  We employ holographic renormalization to show that the variational principle is well-defined. The corresponding Brown--York stress tensor is finite, traceless and conserved. Finally we address possibilities to eliminate the instability and prospects for chiral gravity.
}

\keywords{Cosmological topologically massive gravity, chiral gravity, gravity in three dimensions, logarithmic CFT, holographic renormalization, AdS/CFT}

\preprint{MIT-CTP 3949\\UUITP-08/08}

\begin{document}

\section{Introduction}

Gravity in three dimensions is simple enough to be studied in great depth and complicated enough to make such studies interesting. Pure Einstein--Hilbert gravity exhibits no propagating physical degrees of freedom \cite{Weinberg:1972,Deser:1984tn,Deser:1984dr}. If the theory is deformed by a negative cosmological constant it has black hole solutions \cite{Banados:1992wn}. Another possible deformation is to add a gravitational Chern--Simons term. The resulting theory is called topologically massive gravity (TMG) and, remarkably, contains a massive graviton \cite{Deser:1982vy}. 
Including both terms yields cosmological topologically massive gravity \cite{Deser:1982sv} (CTMG), a theory that exhibits both gravitons and black holes.

Recently, Li, Song and Strominger \cite{Li:2008dq} considered CTMG with the following action	
\begin{equation}
I_{\rm CTMG} 
= \frac{1}{16\pi G}\int_\MM \!\!\!d^3x\sqrt{-g}\,\Big[R+\frac{2}{\ell^2} 
+\frac{1}{2\mu} \,\eps^{\la\mu\nu}\Ga^\rho{}_{\la\si}\,\big(\partial_\mu \Ga^\si{}_{\nu\rho}+\frac23 \,\Ga^\si{}_{\mu\tau}\Ga^\tau{}_{\nu\rho}\big)\Big] \,,
\label{eq:cg1}
\end{equation}
where the negative cosmological constant is parameterized by $\Lambda=-1/\ell^2$.
Notably, the sign of the Einstein--Hilbert action in \eqref{eq:cg1} differs from the choice in \cite{Deser:1982vy} 
that is required to make the graviton energy positive.
The chosen sign in \eqref{eq:cg1} has the advantage of making the BTZ black hole energy positive in the limit of large $\mu$, which is not the case otherwise.

Exploiting the properties of the underlying $SL(2,\mathbb{R})_L\times SL(2,\mathbb{R})_R$ isometry algebra, \cite{Li:2008dq} argued that the would-be-negative energy of the massive graviton mode actually 
is zero if the constants $\mu$ and $\ell$ satisfy the chiral condition\footnote{
The point $\mu\ell = 1$ is special because one of the central charges of the dual boundary CFT vanishes, $c_L=0$, $c_R\neq 0$, and the mass $M$ and angular momentum $J$ of the BTZ black hole solutions satisfy $J = M\ell$. In \cite{Li:2008dq} the theory \eqref{eq:cg1} with \eqref{eq:cg2} was dubbed ``chiral gravity'', assuming that all solutions obey the Brown--Henneaux boundary conditions \cite{Brown:1986nw}. We slightly relax the latter assumption in our discussion, so to avoid confusion we stick to the name ``cosmological topologically massive gravity at the chiral point'' and abbreviate it by CCTMG, where the first C stands for ``chiral''.
}
\eq{
\mu\ell = 1\,.
}{eq:cg2}
Thus, the sign choice in \eqref{eq:cg1} would be admissible as long as \eqref{eq:cg2} holds. For this tuning the massive graviton mode $\psi^M(\mu\ell)$ becomes identical to a mode that exists already in cosmological Einstein gravity. This `left-moving' mode $\psi^L$ is not a physical bulk degree of freedom, and thus the theory appears to lose one physical degree of freedom at the chiral point.

A recent work by Carlip, Deser, Waldron and Wise disputes the claim that no negative energy bulk mode arises for cosmological topologically massive gravity at the chiral point (CCTMG) \cite{Carlip:2008jk}: They find no loss of degree of freedom at the chiral point. The approach of Carlip et al.~is quite different though, which makes a direct comparison cumbersome. 

We clarify here this discrepancy by constructing a negative energy bulk mode that was not considered in \cite{Li:2008dq}, employing their approach. The reason for its existence is the very reason why CCTMG seemingly loses a degree of freedom: When two linearly independent solutions to a differential equation degenerate, a logarithmic solution appears. 
In the present case, the wave function $\psi^M$ of the massive mode degenerates with the left-moving mode $\psi^L$. Therefore, a new solution appears, whose wave-function is given by 
\eq{
\psi^{\rm new} = \lim_{\mu\ell\to 1} \frac{\psi^M(\mu\ell)-\psi^L}{\mu\ell-1}\,.
}{eq:intro}
In this work we study this mode and reveal several intriguing features. In particular it grows linearly in time and the radial coordinate of AdS$_3$. We compute its energy and show that it is finite, negative and time-independent. We also demonstrate that the variational principle is well-defined, including boundary issues. The new mode \eqref{eq:intro} turns out not to contribute to the boundary stress tensor, which is finite, traceless and conserved. To achieve these results we have to extend the analysis of Kraus and Larsen \cite{Kraus:2005zm}, who dropped a term in the Fefferman--Graham expansion that becomes relevant here. Furthermore, we demonstrate that the $L_0$ and $\bar{L}_0$ isometry generators have matrix representations identical to those in logarithmic CFT (LCFT), and therefore the theory is not unitary.

With well-defined variational principle and finite energy, we see no reason to dismiss this mode a priori. Its negative 
energy renders CCTMG unstable, concurrent with \cite{Carlip:2008jk}. 
We find it noteworthy, however, that the destabilizing mode of CCTMG has characteristics quite different from the 
corresponding modes for general $\mu \ell$. For instance, the new mode does not obey the original Brown--Henneaux boundary conditions \cite{Brown:1986nw} (for a 
very recent treatment of CTMG imposing Brown--Henneaux boundary conditions, see \cite{Hotta:2008yq}), but a slightly weaker version thereof that is still consistent with spacetime being asymptotically AdS$_3$. Namely, our Fefferman-Graham expansion for the metric in Gaussian coordinates is of the form
\eq{
ds^2 = \ell^2\, d\rho^2 + \big(e^{2\rho}\,\ga^{(0)}_{ij}+\rho\,\ga^{(1)}_{ij}+\ga^{(2)}_{ij}+\dots\big)\,dx^i dx^j\,,
}{eq:angelinajolie}
which reduces to the Brown--Henneaux case for vanishing $\ga^{(1)}$ only.
Moreover, the new mode is not periodic in time and therefore does not contribute to a finite temperature partition function. This could mean that it is nevertheless possible to make sense of CTMG exactly at the chiral point, as conjectured by Li, Song and Strominger \cite{Li:2008dq}. This would have to involve a consistent truncation of the new mode.
We shall argue in the Conclusions that even without such a truncation CCTMG and its related LCFT provide interesting subjects for further studies.

This paper is organized as follows. We begin in Section \ref{se:2} by recalling basic features of CTMG and CCTMG. We construct the new physical mode and calculate its energy in Section \ref{se:3}. In Section \ref{se:4} we show that this mode is a valid classical solution (including boundary issues) and we calculate the boundary stress tensor. We conclude with a brief summary and discussion of future prospects for CCTMG and chiral gravity in Section \ref{se:5}.

Before starting, we mention some of our conventions. We set $16\pi G=1$ and otherwise use the same conventions for signature and sign definitions\footnote{
The Chern-Simons term in \eqref{eq:cg1} has a sign different from \cite{Li:2008dq}, thus correcting a typo in that work.
} 
as in \cite{Li:2008dq}, including Riemann tensor $R^\mu{}_{\nu\si\la}=\partial_\si\Ga^\mu{}_{\nu\la}+\dots$, Ricci tensor $R_{\mu\nu}=R^\la{}_{\mu\la\nu}$ and epsilon symbol $\epsilon^{012}=\epsilon^{01}=+1$. The epsilon-tensor is denoted by $\eps^{\la\mu\nu}=\epsilon^{\la\mu\nu}/\sqrt{-g}$. For sake of specificity we consider exclusively $\ell>0$. We use Greek indices for 3-dimensional tensors and Latin indices for 2-dimensional ones. For adapted coordinates we take $x^0=\tau$, $x^1=\phi$ and $x^2=\rho$. Our conventions for boundary quantities and the Fefferman--Graham expansion are summarized in Appendix \ref{app:A}.

\section{CTMG and CCTMG} \label{se:2}

In this section we review the powerful formulation of linearized CTMG developed 
in \cite{Li:2008dq}. We put particular emphasis on the behavior at the chiral point $\mu\ell = 1$.

The background metric $\bar{g}_{\mu\nu}$ is that of global AdS$_3$,
\eq{
ds^2 = \bar{g}_{\mu\nu}\,dx^\mu dx^\nu= \ell^2\big(-\cosh^2{\!\!\rho}\, d\tau^2 +\sinh^2{\!\!\rho}\,d\phi^2+d\rho^2\big)\,,
}{eq:cg20}
whose isometry group is $SL(2,\mathbb{R})_L\times SL(2,\mathbb{R})_R$. In light-cone coordinates $u=\tau+\phi$, $v=\tau-\phi$ the $SL(2,\mathbb{R})_L$ generators read
\begin{align}
L_0 &= i\partial_u \label{eq:cg21}\\
L_{-1} &= ie^{-iu}\Big[\frac{\cosh{2\rho}}{\sinh{2\rho}}\partial_u-\frac{1}{\sinh{2\rho}}\partial_v + \frac i2 \partial_\rho\Big]\label{eq:cg22}\\
L_1 &= ie^{iu}\Big[\frac{\cosh{2\rho}}{\sinh{2\rho}}\partial_u-\frac{1}{\sinh{2\rho}}\partial_v - \frac i2 \partial_\rho\Big] \label{eq:cg23}
\end{align}
with algebra
\eq{
\big[L_0,L_{\pm 1}\big]=\mp L_{\pm 1}\,,\qquad \big[L_1,L_{-1}\big]=2L_0
}{eq:cg24}
and quadratic Casimir
\eq{
L^2=\frac12 \,\big(L_1 L_{-1} + L_{-1} L_1\big)-L_0^2\,.
}{eq:cg25}
The $SL(2,\mathbb{R})_R$ generators $\bar L_0$, $\bar L_{-1}$, $\bar L_1$ satisfy the same algebra and are given by \eqref{eq:cg21}-\eqref{eq:cg23} with $u\leftrightarrow v$ and $L\leftrightarrow \bar L$.

The full non-linear equations of motion of CTMG read
\eq{
G_{\mu\nu} + \frac{1}{\mu}\, C_{\mu\nu} = 0\,,
}{eq:cg51}
where 
\eq{
G_{\mu\nu} = R_{\mu\nu} - \frac12\, g_{\mu\nu} R -\frac{1}{\ell^2}\, g_{\mu\nu}
}{eq:cg45}
is the Einstein tensor (including cosmological constant) and 
\eq{
C_{\mu\nu} = \frac12\,\eps_\mu{}^{\al\be} \,\nabla_\al R_{\be\nu} + (\mu\leftrightarrow\nu)
}{eq:cg58}
is essentially the Cotton tensor. To look for 
perturbative solutions to \eqref{eq:cg51}, we write the metric
as the sum of the AdS$_3$ background \eqref{eq:cg20} and fluctuations $h_{\mu\nu}$.
\eq{
g_{\mu\nu} = \bar{g}_{\mu\nu} + h_{\mu\nu}.	
}{eq:cg34}
Expanding in $h_{\mu\nu}$ produces the linearized equations of motion
\eq{
G_{\mu\nu}^{\rm lin} + \frac{1}{\mu}\, C_{\mu\nu}^{\rm lin} = 0\,,
}{eq:cg52}
where 
\eq{
G_{\mu\nu}^{\rm lin}=R_{\mu\nu}^{\rm lin}-\frac12\,\bar{g}_{\mu\nu}R^{\rm lin}+\frac{2}{\ell^2}\,h_{\mu\nu}
}{eq:cg49}
and
\eq{
C_{\mu\nu}^{\rm lin} = \frac12\, \eps_\mu{}^{\al\be}\bar\nabla_\al G_{\be\nu}^{\rm lin} + (\mu\leftrightarrow\nu)
}{eq:cg50}
are the linear versions of the Einstein and Cotton tensors, respectively. Expressions for the linearized Ricci tensor $R^{\rm lin}_{\mu\nu}$ and Ricci scalar $R^{\rm lin}$ can be found in \cite{Deser:2003vh}.

By choosing the transverse and traceless gauge
\eq{
\bar\nabla_\mu h^{\mu\nu}=0\,,\qquad \bar g^{\mu\nu} h_{\mu\nu}=0
}{eq:cg35}
the linearized equations of motion \eqref{eq:cg52} take the form
\eq{
\big({\mathcal D}^R{\mathcal D}^L{\mathcal D}^M h\big)_{\mu\nu}=0\,.
}{eq:cg40}
The mutually commuting differential operators ${\mathcal D}^{L/R/M}$ are given by
\eq{
({\mathcal D}^{L/R})_\mu{}^\nu = \de_\mu^\nu \pm \ell \eps_\mu{}^{\al\nu}\,\bar\nabla_\al\,,\qquad ({\mathcal D}^M)_\mu{}^\nu = \de_\mu^\nu + \frac1\mu\, \eps_\mu{}^{\al\nu}\,\bar\nabla_\al\,.
}{eq:cg36}
Notice that for CCTMG ${\mathcal D}^{M} = {\mathcal D}^{L}$, and that the equations of motion 
for this case read
\eq{
\big({\mathcal D}^R{\mathcal D}^L{\mathcal D}^L h\big)_{\mu\nu}=0\,.
}{eq:cg37}
For generic values of $\mu$ and $\ell$ the three linearly independent solutions to \eqref{eq:cg40} can be taken to satisfy
\eq{
\big({\mathcal D}^L h^L\big)_{\mu\nu}=0\,,\qquad \big({\mathcal D}^R h^R\big)_{\mu\nu}=0\,,\qquad \big({\mathcal D}^M h^M\big)_{\mu\nu}=0\,.
}{eq:cg38}
These branches of solutions are referred to as left-moving, right-moving and massive gravitons, respectively.
Solely the latter entails physical bulk degrees of freedom.
The basis of solutions \eqref{eq:cg38} becomes 
inadequate at the chiral point $\mu\ell = 1$, since, at that point, the $L$ and $M$ branches coincide. 
In the next Section we remedy this deficiency by explicitly constructing a mode $h^{\rm new}_{\mu\nu}$ satisfying\footnote{
We are grateful to Roman Jackiw for suggesting to perform such a construction.
}
\eq{
\big({\mathcal D}^L{\mathcal D}^L h^{\rm new}\big)_{\mu\nu}=0\,,\qquad \big({\mathcal D}^L h^{\rm new}\big)_{\mu\nu}\neq 0\,.
}{eq:cg39}
Using the $SL(2,\mathbb{R})$ algebra, \cite{Li:2008dq} finds all solutions to \eqref{eq:cg38}. These sets of solutions consist of
primaries satisfying $L_1 \psi = \bar{L}_1 \psi = 0$, and descendants obtained by acting with $L_{-1}$ and $\bar{L}_{-1}$.
The explicit form of the wave functions for the massive and left-moving primaries will be of importance to us, so we recall them here.
\eq{
\psi^M_{\mu\nu} = e^{-(3/2 + \mu\ell/2)iu - (-1/2 + \mu\ell/2)iv}\frac{\sinh^2{\!\!\rho}}{(\cosh{\rho})^{1+\mu\ell}} \left(\begin{array}{ccc}
1 & 1 & i a \\
1 & 1 & i a \\
ia & ia & -a^2 
\end{array}\right)_{\!\!\!\mu\nu} 
}{eq:cg46}
where
\eq{
a:=\frac{1}{\sinh{\rho}\,\cosh{\rho}}
}{eq:cg47}
The left-mover $\psi^L_{\mu\nu}$ is obtained from \eqref{eq:cg46} by setting $\mu\ell = 1$. The real and imaginary
parts of $\psi_{\mu\nu}$ separately solve the equations of motion. We take
\eq{
h_{\mu\nu}=\Re\,{\psi_{\mu\nu}}\,.
}{eq:cg48}
This concludes our recapitulation of CTMG.

\section{Logarithmic mode with negative energy} \label{se:3}

In this Section we construct and discuss the new mode of CCTMG. Using the explicit form \eqref{eq:cg46} of $\psi^M$ it is straight-forward to perform the standard construction:
\eq{
\psi^{\rm new}_{\mu\nu}:=\lim_{\mu\ell\to 1}\frac{\psi^M_{\mu\nu}(\mu\ell)-\psi^L_{\mu\nu}}{\mu\ell - 1} = y(\tau,\rho) \,\psi^L_{\mu\nu}
}{eq:cg26}
where we define the function $y$ by
\eq{
y(\tau, \rho):=-i\tau-\ln{\cosh{\rho}}\,.
}{eq:cg27}
When analyzing the asymptotics of the new mode it will be convenient to have an explicit expression for $h^{\rm new}_{\rm \mu\nu}$. Using \eqref{eq:cg46}-\eqref{eq:cg48} we obtain 
\begin{align}
\;\;\; h^{\rm new}_{\rm \mu\nu} = \frac{\sinh{\rho}}{\cosh^3\!\!{\rho}}\,\big(\cos{(2u)}\,\tau-\sin{(2u)}\,\ln{\cosh{\rho}}\big)&\left( \begin{array}{ccc}
0 & 0 & 1 \\
0 & 0 & 1 \\
1 & 1 & 0
\end{array} \right)_{\!\!\!\mu\nu} \nonumber \\
-\tanh^2\!\!{\rho}\,\big(\sin{(2u)}\,\tau+\cos{(2u)}\,\ln{\cosh{\rho}}\big)&\left( \begin{array}{ccl}
1 & 1 & 0 \\
1 & 1 & 0 \\
0 & 0 & -\sinh^{-2}\!\!{\rho}\,\cosh^{-2}\!\!{\rho}
\end{array} \right)_{\!\!\!\mu\nu}.
\label{eq:cg83}
\end{align}
We see that the new mode grows linearly in time, and also (asymptotically) in the radial coordinate $\rho$. 

To show that $\psi^{\rm new}_{\mu\nu}$ solves the bulk equations of motion, let us determine the action of the isometry algebra.
Acting on $y$ we obtain
\eq{
L_0\,y=\bar L_0\,y = \frac12\,,\qquad L_1\,y = \bar{L}_1\,y = 0\,.
}{eq:cg28}
Correspondingly, on $\psi^{\rm new}$ the action is
\eq{
L_0\,\psi^{\rm new}_{\mu\nu} = 2 \psi^{\rm new}_{\mu\nu}+\frac12\,\psi^L_{\mu\nu}\,, \qquad \bar L_0\,\psi^{\rm new}_{\mu\nu} = \frac12\,\psi^L_{\mu\nu}\,, \qquad L_1\,\psi^{\rm new} = \bar{L}_1\,\psi^{\rm new} = 0\,.
}{eq:cg31}
Note that $\psi^{\rm new}$ is not an eigenstate of $L_0$ or $\bar{L}_0$, but only of $L_0-\bar L_0$. Because of the relations \eqref{eq:cg31} it is impossible to decompose $\psi^{\rm new}$ as a linear combination of eigenstates to $L_0$ and $\bar L_0$. The representation of $L_0$ and $\bar L_0$ as matrices, 
\eq{ L_0 \left(\begin{array}{c} \psi^{\rm new} \\ \psi^L
\end{array}\right) = \left(\begin{array}{cc}
2 & \frac12 \\
0 & 2
\end{array}\right) \left(\begin{array}{c} \psi^{\rm new} \\ \psi^L \end{array}\right)\,,\qquad \bar L_0 \left(\begin{array}{c} \psi^{\rm new} \\ \psi^L
\end{array}\right) = \left(\begin{array}{cc} 0 & \frac12 \\ 0 & 0
\end{array}\right) \left(\begin{array}{c} \psi^{\rm new} \\ \psi^L \end{array}\right)\,, 
}{eq:cg79} 
shows that their Jordan normal form is the same as in LCFT \cite{Gurarie:1993xq}. In the parlance of LCFT literature $\psi^{\rm new}$ is the logarithmic partner of $\psi^L$.
(For reviews see \cite{Flohr:2001zs,Gaberdiel:2001tr}; for some applications to AdS/LCFT see \cite{Ghezelbash:1998rj,Myung:1999nd,Kogan:1999bn,Lewis:1999qv, MoghimiAraghi:2004ds}.\footnote{The relation to LCFTs was pointed out by John McGreevy during a talk by Andy Strominger at MIT. We thank John McGreevy for discussions on LCFTs.}) 

From the equations \eqref{eq:cg31} we deduce
\eq{
({\mathcal D}^R{\mathcal D}^L \psi^{\rm new})_{\mu\nu} = -\ell^2\, \big(\bar\nabla^2+\frac{2}{\ell^2}\big)\psi^{\rm new}_{\mu\nu} = 2\,\big(L^2+\bar L^2+2\big)\psi^{\rm new}_{\mu\nu}=-2\psi^L_{\mu\nu}
}{eq:cg32}
and consequently
\eq{
\big({\mathcal D}^L {\mathcal D}^R {\mathcal D}^L \psi^{\rm new}\big)_{\mu\nu}=0\,.
}{eq:cg33}
The identity \eqref{eq:cg33} shows that $\psi^{\rm new}$ solves the classical equations of motion.
Acting on $\psi^{\rm new}$ with $L_{-1}$ and $\bar{L}_{-1}$ produces a tower of descendants.

As expected on general grounds, the new mode $\psi^{\rm new}$ is indeed a physical mode and not just pure gauge. To prove this it is sufficient to demonstrate that there is no gauge preserving coordinate transformation $\xi_{\mu}$ that annihilates $\psi^{\rm new}_{\mu\nu}$,
\eq{
\bar \nabla_{(\mu}\xi_{\nu)} + \psi^{\rm new}_{\mu\nu} = 0\,.
}{eq:cg53}
The quickest way to show that \eqref{eq:cg53} has no solution for $\xi_\mu$ is as follows: for any $\xi_{\mu}$ preserving the gauge conditions $\nabla_{(\mu}\xi_{\nu)}$ solves
the linearized Einstein equations,\footnote{We thank Wei Li and Wei Song for providing this argument.} while $\psi^{\rm new}_{\mu\nu}$ does not. We also mention that despite of the linear divergence of $\psi^{\rm new}$ in the radial coordinate $\rho$ the linearized approximation does not break down asymptotically, i.e., \eqref{eq:cg83} really is a small perturbation of the AdS$_3$ background.

Let us now compute the energy of the new mode. We do this by the procedure described in \cite{Buchbinder:1992pe,Li:2008dq}.
The Hamiltonian is given by
\eq{
H = \int dx^2 \,\big(\dot{h}_{\mu\nu} \Pi^{(1)\mu\nu} + (\bar{\nabla}_0 \dot{h}_{\mu\nu}) \Pi^{(2)\mu\nu} - {\mathcal L} \big)\,,
}{eq:cg54}
where ${\mathcal L}$ is the Lagrange density expanded to quadratic order in $h$,
and the canonical momenta $\Pi^{(1)\mu\nu}$ and $\Pi^{(2)\mu\nu}$ are given by
\eq{
\Pi^{(1)\mu\nu} = -\frac{\sqrt{-\bar{g}}}{4}\Big(\bar{\nabla}^0 (2h^{\mu\nu} + \ell\,\eps^{\mu\alpha}{}_{\be}\bar{\nabla}_{\alpha}h^{\be\nu} ) - \ell\,\eps_{\be}{}^{0\mu}(\bar{\nabla}^2 + \frac{2}{\ell^2})h^{\be\nu}  \Big)
}{eq:cg55}
\eq{
\Pi^{(2)\mu\nu} = -\frac{\sqrt{-\bar{g}} \,\bar{g}^{00}}{4} \,\ell\,\eps_{\be}{}^{\lambda\mu}\bar{\nabla}_{\lambda} h^{\be\nu}.
}{eq:cg56}
It is slightly lengthy, but straightforward to evaluate \eqref{eq:cg54} on the solution \eqref{eq:cg83}. By virtue of the on-shell relations \eqref{eq:cg32} and ${\mathcal L}=0$ the on-shell Hamiltonian reduces to
\begin{multline}
H\big|_{\rm EOM} = \frac12 \int d^2x\sqrt{-\bar g}\,\Big[\big((\bar\nabla^0\dot h^{\rm new}_{\mu\nu})(h^{\mu\nu}_{\rm new}+\ell\,\eps^{\mu\al}{}_\be \bar\nabla_\al h_{\rm new}^{\be\nu})+\frac1\ell\,\dot h_{\mu\nu}^{\rm new}\,\eps^{0\mu}{}_\be\,h^{\be\nu}_{L}\big) \\
-\bar g^{00}\partial_0 \big(\dot{h}_{\mu\nu}^{\rm new}(h^{\mu\nu}_{\rm new}+\frac{\ell}{2}\,\eps^{\mu\al}{}_\be\bar\nabla_\al h^{\be\nu}_{\rm new})\big)\Big]=: E^{\rm new}
\label{eq:cg78}
\end{multline}
Note the appearance of both $h^{\rm new}$ and $h^L$ in the integrand. Evaluating the integral \eqref{eq:cg78} leads to the result (with $16\pi G$ reinserted)
\eq{
E^{\rm new} = \frac{2\pi}{16\pi G\,\ell^3} \int\limits_1^\infty dx\, \Big( \frac{8}{x^9}\,\log{x}-\frac{9}{2x^9} - \frac{2}{x^7}\,\log{x} +\frac{1}{x^7} \Big) = - \frac{47}{1152G\,\ell^3}\,.
}{eq:cg57}
We see explicitly that the energy is finite, negative and time-independent. While the finiteness of \eqref{eq:cg57} may seem surprising considering that $h^{\rm new}$ diverges, we recall that it is not unusual for a mode to grow linear in time and still have time independent finite energy. Comparable precedents are free motion in Newtonian mechanics and static spherically symmetric solutions of the Einstein-massless-Klein-Gordon model with a scalar field that grows linearly in time \cite{Wyman:1981bd}.

We conclude that the new mode \eqref{eq:cg83} for CCTMG cannot be dismissed on physical grounds, since it is not merely pure gauge and its energy remains bounded. Moreover, its energy is negative and thus CCTMG is unstable. The boundary issues considered in the next Section do not alter this conclusion.

\section{Variational principle and boundary stress tensor} \label{se:4}

We pose now the relevant question whether the new mode \eqref{eq:cg83} is actually a classical solution of CCTMG. To this end not only the bulk equations of motion \eqref{eq:cg37} must hold, as they do indeed, but also all boundary terms must be canceled so that the first variation of the on-shell action
\begin{multline}
\de I_{\rm CG}\big|_{\rm EOM} = -\int_{\dM}\!\!\!\!d^2x \sqrt{-\ga}\,\Big(K^{ij}-\big(K-\frac1\ell\big)\ga^{ij}\Big)\de\ga_{ij}\\
+\ell\int_{\dM}\!\!\!\!d^2x\,\epsilon^{ij}\,\Big(-R^{k\rho}{}_{j\rho}\,\de\ga_{ik}+K_i{}^k\,\de K_{kj}-\frac12 \Ga^k{}_{li}\,\de\Ga^l{}_{kj}\Big)
\label{eq:cg9}
\end{multline}
vanishes for all variations preserving the boundary conditions. While answering this question in the affirmative, we shall obtain as a byproduct the result for the boundary stress tensor $T^{ij}$, which follows also from the variation of the on-shell action
\eq{
\de I_{\rm CCTMG}\big|_{\rm EOM} = \frac12 \int_{\dM}\!\!\!\!d^2x\sqrt{-\ga^{(0)}}\,T^{ij}\,\de\ga_{ij}^{(0)}\,.
}{eq:cg10}
Here $\ga^{(0)}$ is the metric on the conformal boundary as defined in Appendix \ref{app:A}.
In order to proceed we must supplement the bulk action \eqref{eq:cg1} with appropriate boundary terms.

CCTMG requires two kinds of boundary terms, as most other gravitational theories do: a Gibbons--Hawking--York boundary term for making the Dirichlet boundary value problem well-defined, and a boundary counterterm for making the variational principle well-defined. It was shown by Kraus and Larsen \cite{Kraus:2005zm} (for related considerations see also \cite{Solodukhin:2005ah}) that the fully supplemented CCTMG action is given by
\begin{multline}
I_{\rm CCTMG} = \int_\MM \!\!\!d^3x\sqrt{-g}\,\Big(R+\frac{2}{\ell^2}\Big) + 2\int_{\dM}\!\!\!\!d^2x\sqrt{-\ga}\,\Big(K-\frac1\ell\Big)\\
+\frac\ell2\int_\MM \!\!\!d^3x\sqrt{-g}\,\eps^{\la\mu\nu}\Ga^\rho{}_{\la\si}\Big(\partial_\mu \Ga^\si{}_{\nu\rho}+\frac23\, \Ga^\si{}_{\mu\tau}\Ga^\tau{}_{\nu\rho}\Big)
\label{eq:cg3}
\end{multline}
Its first variation leads to \eqref{eq:cg9} above.
Remarkably, the boundary terms are just the ones that are present already in cosmological Einstein-Hilbert gravity, i.e., the terms in the first line of \eqref{eq:cg3}. However, the result \eqref{eq:cg3} was derived assuming a restricted Fefferman-Graham expansion of the boundary metric, i.e., one that does not involve the term linear in $\rho$ in \eqref{eq:cg4} below. This is not sufficient to encompass the new mode described in Section \ref{se:3}. Rather, we get the expansion announced in \eqref{eq:angelinajolie}, viz.
\eq{
\ga_{ij}=e^{2\rho}\ga^{(0)}_{ij} + \rho\, \ga_{ij}^{(1)} + \ga_{ij}^{(2)} + \dots
}{eq:cg4}
for the boundary metric, which coincides with  \cite{Kraus:2005zm} for $\ga_{ij}^{(1)}=0$ only. Here $\ga^{(0)}_{ij}$ is the conformal metric at the boundary, 
$\ga^{(1)}_{ij}$
describes the linearly growing contribution and
$\ga_{ij}^{(2)}$
the constant contribution. 

Let us comment briefly on the linear term in \eqref{eq:cg4}. Such a term is always present in pure gravity for odd-dimensional AdS spacetimes with dimension $D \geq 5$. In $D=3$ the coefficient in front of this term is set to zero by the Einstein equations \cite{deHaro:2000xn}, and it is not included in the boundary conditions of Brown and Henneaux\cite{Brown:1986nw}. However, it is also well-known that violations of the original Brown--Henneaux boundary conditions can arise even in three dimensions if gravity couples to matter \cite{Henneaux:2002wm}, 
and that the linear term in \eqref{eq:cg4} does not spoil the property of spacetime being asymptotically AdS \cite{Skenderis:2002wp}. Interestingly, the coupling to a Chern-Simons term leads to such a linear term, as we demonstrate here explicitly.  

To identify the coefficients $\gamma^{(i)}$, we recall that the full metric is given by \eqref{eq:cg34}, where $\bar{g}_{\mu\nu}$ is the background metric \eqref{eq:cg20} and $h_{\mu\nu} = h^{\rm new}_{\mu\nu}$ is the new mode \eqref{eq:cg83}. The boundary metric 
\eq{
\ga^{(0)}_{ij} = \frac{\ell^2}{4} \left(\begin{array}{rr}
-1 & 0 \\
0 & 1 
\end{array}\right)_{\!ij}
}{eq:cg43}
is (trivially) conformal to the Minkowski metric, the linearly growing contribution reads
\eq{
\ga^{(1)}_{ij}=-\cos{(2u)}\,\left(\begin{array}{cc}
1 & 1 \\
1 & 1 
\end{array}\right)_{\!ij}
}{eq:cg5}
and the constant contribution is given by
\eq{
\ga^{(2)}_{ij} = -\big(\sin{(2u)}\,\tau-\cos{(2u)}\ln{2}\big) \,\left(\begin{array}{cc}
1 & 1 \\
1 & 1 
\end{array}\right)_{\!ij} -\frac{\ell^2}{2}\,\left(\begin{array}{cc}
1 & 0 \\
0 & 1 
\end{array}\right)_{\!ij}\,.
}{eq:cg6}
The first term in \eqref{eq:cg6} comes from $h^{\rm new}$ and the second one from the next-to-leading order term of the AdS$_3$ background.
As explained in Appendix \ref{app:A} we use \eqref{eq:cg4} to expand relevant quantities like extrinsic curvature for large $\rho$.

Variations that preserve the boundary conditions are those where $\de\ga^{(0)}$ vanishes, but $\de\ga^{(1)}$ and $\de\ga^{(2)}$ may be finite. Thus, a well-defined variational principle requires that only $\de\ga^{(0)}$ remains in \eqref{eq:cg9} after taking the limit $\rho\to\infty$. We have checked that all the terms appearing in \eqref{eq:cg9} indeed contain exclusively $\de\ga^{(0)}$-terms, see \eqref{eq:ap9}-\eqref{eq:ap13} in Appendix \ref{app:A}. Therefore, we have generalized the conclusions of Kraus and Larsen that CTMG has a well-defined variational principle to the case where the Fefferman-Graham expansion \eqref{eq:cg4} has a non-vanishing contribution from \eqref{eq:cg5}.

From the result \eqref{eq:ap13} in Appendix \ref{app:A} we can now read off the boundary stress tensor as defined in \eqref{eq:cg10}.
\eq{
T^{ij} = \lim_{\rho\to\infty} \frac1\ell \,\Big[\rho\,\big(\ga^{ij}_{(1)}-\ga^{il}_{(1)}\ga_{lk}^{(0)}\eps^{kj}\big) -\frac12\,\big(\ga^{ij}_{(1)}-3\ga^{il}_{(1)}\ga_{lk}^{(0)}\eps^{kj}\big) + \ga_{(2)}^{ij} - \ga^{il}_{(2)}\ga_{lk}^{(0)}\eps^{kj} \Big] +(i\leftrightarrow j)
}{eq:cg11}
For vanishing $\ga^{(1)}$ this coincides with the result\footnote{We note that in \cite{Kraus:2005zm} there is a sign change between appendix and body of the paper.} (5.14) of \cite{Kraus:2005zm} if we take into account the tracelessness of $\ga^{(2)}$. 
For non-vanishing $\ga^{(1)}$ apparently the boundary stress tensor \eqref{eq:cg11} diverges. However, with \eqref{eq:cg5} we see that the expression
\eq{
\ga^{ij}_{(1)} - \ga^{il}_{(1)}\ga_{lk}^{(0)}\eps^{kj} = 0
}{eq:cg12}
actually vanishes identically. Therefore, the linear divergence in $\rho$ is not present in the boundary stress tensor for the mode \eqref{eq:cg83}. 
The equations \eqref{eq:cg43}-\eqref{eq:cg12} establish our result for the boundary stress tensor (with $16\pi G$ reinserted)
\eq{
T^{ij} = -\frac{1}{\pi G\,\ell^3} \, \left(\begin{array}{cc}
1 & 1 \\
1 & 1 
\end{array}\right)^{\!ij} - \frac{2}{\pi G\,\ell^5} \,\cos{(2u)} \, \left(\begin{array}{rr}
1 & -1 \\
-1 & 1 
\end{array}\right)^{\!ij}\,.
}{eq:cg14}
The AdS stress tensor is interpreted as the Casimir energy of the dual field theory \cite{Balasubramanian:1999re,Emparan:1999pm}. 
The boundary stress tensor \eqref{eq:cg14} is finite, traceless and conserved. 
Except for the crucial first property of finiteness, these features might have been anticipated on general grounds. The finiteness confirms our conclusion of the previous Section: The new mode \eqref{eq:cg83} cannot be dismissed on physical grounds.  

\section{Conclusions} \label{se:5}

To summarize, we have investigated CCTMG \eqref{eq:cg1}, \eqref{eq:cg2} at the linearized level along the lines of \cite{Li:2008dq} and found a new mode \eqref{eq:cg83}, concurrent with the analysis of \cite{Carlip:2008jk}. We checked that this mode is physical, i.e., not pure gauge, and that it has finite, time-independent negative energy \eqref{eq:cg57}. We showed also that this mode is a valid classical solution in the sense that the variational principle is well-defined. Furthermore we demonstrated that it has a Fefferman-Graham expansion \eqref{eq:cg4} and therefore does not spoil the property of spacetime being asymptotically AdS$_3$. Thus, we may conclude that CCTMG is unstable, because the new mode is physically acceptable, but has negative energy. As a byproduct we calculated the boundary stress tensor \eqref{eq:cg14} and found that it is finite, traceless and conserved. By analyzing the action of the isometry algebra on the new mode, we concluded that a dual CFT describing this mode must be a logarithmic CFT and therefore is not unitary.

While the analysis in the current work used the linearized approximation, the new mode is present also non-perturbatively. This can be checked easily by a canonical analysis, which reveals that nothing special happens with the dimension of the physical phase space as the chiral point \eqref{eq:cg2} is approached.\footnote{We thank Steve Carlip for conveying this information to us. We have convinced ourselves independently that this statement is true, but we do not include the corresponding analysis here. We just mention that a simple way to derive this statement is to exploit the first order formulation of \cite{Baekler:1991} and count the number of first- and second-class constraints. See also Ref.~\cite{Park:2008yy} for a recent canonical analysis along these lines and Ref.~\cite{Deser:1991qk} for a corresponding analysis at $\ell=0$.}

It is conceivable that nonperturbative effects stabilize CCTMG, i.e., that the instability is an artifact of perturbation theory, but we have found no evidence for this suggestion. Since very few exact solutions of CTMG are known \cite{Percacci:1986ja,Hall:1987bz,Nutku:1993eb,Aliev:1996eh,Dereli:2000fm,Bouchareb:2007yx} and because the new mode \eqref{eq:cg83} exhibits two commuting Killing vectors, a reasonable strategy to find relevant nonperturbative solutions would be the consideration of exact solutions with two commuting Killing vectors. To this end a 2-dimensional dilaton gravity \cite{Grumiller:2002nm} 
approach extending the analysis of \cite{Guralnik:2003we,Grumiller:2003ad} could be helpful (see also \cite{2d}). We also recall that the gravitational modes have positive energy --- not just for CCTMG, but generically --- if the sign of the Einstein-Hilbert term in \eqref{eq:cg1} is reversed. This sign change, however, leads to negative energy for BTZ black hole solutions\footnote{In \cite{Carlip:2008jk} it was pointed out that this issue is resolved if one finds a superselection sector in which BTZ black holes are excluded.} as emphasized in \cite{Li:2008dq}.

CCTMG can exist as a meaningful theory, which one might call chiral gravity, if the new mode is absent. Thus, it is of interest to point out applications where the new mode is eliminated. If one imposes boundary conditions that are stricter than required by the variational procedure then the new mode can be discarded. This is the case if one imposes the original Brown--Henneaux boundary conditions \cite{Brown:1986nw}. However, we reiterate that the expansion \eqref{eq:cg4} is consistent with spacetime being asymptotically AdS$_3$ \cite{Skenderis:2002wp}, so a priori there is no reason to impose stronger conditions. Indeed, insisting on this stronger set of boundary conditions would also eliminate physically interesting solutions in similar theories of gravity \cite{Henneaux:2002wm}. 
Whether such a truncation of CCTMG to chiral gravity is quantum mechanically consistent remains as a pivotal open issue.\footnote{We thank Andy Strominger for helpful discussions on these issues.}

Alternatively, if one imposes periodic boundary conditions $\tau = \tau + \beta$ on the metric the new mode \eqref{eq:cg83} is eliminated since it is linear in $\tau$. Therefore, at finite (but arbitrarily small) temperature the new mode appears to become irrelevant. This conclusion applies as well to the descendants, which are obtained by acting with $L_{-1}$ and $\bar{L}_{-1}$ on \eqref{eq:cg26} and therefore have a contribution linear in $\tau$.

The considerations in the previous paragraphs might be of interest for the Euclidean approach/CFT approach pursued in \cite{Witten:2007kt,Manschot:2007zb,Gaiotto:2007xh,Gaberdiel:2007ve,Avramis:2007gx,Yin:2007gv,Yin:2007at,Maloney:2007ud,Giombi:2008vd}. We conclude with three options, all of which are worthwhile pursuing:
\enlargethispage{1cm}
\begin{enumerate}
\item A consistent quantum theory of Euclidean chiral gravity with a chiral CFT dual may exist if the truncation of CCTMG can be shown to be admissible. At the boundary this would involve a truncation of a LCFT to a unitary CFT. One can check the viability of this option by studying correlators like $\langle\psi^{\rm new}\,\psi^L\,\psi^L\rangle$. If they are non-vanishing no truncation is possible. 
\item If a truncation turns out to be impossible then an alternative option is to find a unitary completion of the theory. 
\item Even without truncation or completion CCTMG and its related LCFT provide interesting subjects for further studies.\footnote{
In this case the attribute ``chiral'' in CCTMG is slightly misleading
since the dual LCFT is not chiral even for $c_L=0$. We thank Matthias
Gaberdiel for pointing this out. See also Refs.~\cite{Gaberdiel:1998ps,Eberle:2006zn}.
} LCFTs are not unitary, but still useful as physical models \cite{Flohr:2001zs,Gaberdiel:2001tr}. One could learn something about 3-dimensional gravity in general and about the instability described here in particular, by studying the dual LCFT. On the other hand, studying the bulk theory along the lines of the present work may also shed some light on properties of the dual LCFT, via the dictionary of AdS/LCFT \cite{Ghezelbash:1998rj,Myung:1999nd,Kogan:1999bn,Lewis:1999qv, MoghimiAraghi:2004ds}.
\end{enumerate}

\acknowledgments

We thank Steve Carlip, Stanley Deser, Henriette Elvang, Matthias Gaberdiel, Thomas Hartman, Alfredo Iorio, Roman Jackiw, Per Kraus, Wei Li, Alexander Maloney, John McGreevy, Robert McNees, Wei Song, Andy Strominger, Alessandro Torrielli, Andrew Waldron, Derek Wise and Xi Yin for discussions. NJ thanks the CTP at MIT for its kind hospitality during the main part of this work.

This work is supported in part by funds provided by the U.S. Department of Energy (DoE) under the cooperative research agreement DEFG02-05ER41360. DG is supported by the project MC-OIF 021421 of the European Commission under the Sixth EU Framework Programme for Research and Technological Development (FP6). The research of NJ was supported in part by the STINT CTP-Uppsala exchange program.

\paragraph{Note added:} Previous versions of this paper contained a sign error in (4.1) with relevant consequences for the finite part of the Brown--York stress tensor (4.8) and (4.10). This sign error was corrected in an erratum \cite{erratum}, prepared together with Sabine Ertl. 

\begin{appendix}

\section{Fefferman-Graham expansion} \label{app:A}

While the conclusions of the analysis below are gauge independent, it is convenient to use an adapted coordinate system. Even though the bulk metric \eqref{eq:cg34} is not in Gaussian coordinates
\eq{
  ds^2=g_{\mu\nu}\,dx^\mu dx^\nu=d\hat\rho^2+\ga_{ij} \,dx^i dx^j
}{eq:ap1}
its shift vector and lapse function do not contribute to the relevant order in a large $\hat\rho$ expansion. Thus, for boundary purposes the bulk metric \eqref{eq:cg34} actually is of the form \eqref{eq:ap1}, up to a  factor of $\ell^2$, which we shall take into account in the very end. Therefore, we can exploit the standard features of Gaussian coordinates, e.g.~that the outward pointing unit normal vector $n^\mu$ has only a $\hat\rho$-component, $n^{\hat\rho}=1$, $n^i=0$, and that the first fundamental form $h_{\mu\nu} = g_{\mu\nu}-n_\mu n_\nu$ has only non-vanishing $ij$-components given by $h_{ij}=\ga_{ij}$. Thus, $\ga_{ij}$ is the boundary metric. Similarly, the second fundamental form $K_{\mu\nu}=h_\mu{}^\al h_\nu{}^\be \nabla_\al n_\be$ has only non-vanishing $ij$-components given by $K_{ij}=-\Ga^{\hat\rho}{}_{ij}$.

We expand the boundary metric in the limit of large $\hat\rho$
\eq{
\ga_{ij}=e^{2\hat\rho/\ell}\,\ga^{(0)}_{ij} + \frac{\hat\rho}{\ell}\, \ga_{ij}^{(1)} + \ga_{ij}^{(2)} + \dots
}{eq:ap2}
as well as its inverse,
\eq{
\ga^{ij}=e^{-2\hat\rho/\ell}\,\ga_{(0)}^{ij} - e^{-4\hat\rho/\ell}\,\frac{\hat\rho}{\ell}\, \ga^{ij}_{(1)} - e^{-4\hat\rho/\ell}\,\ga^{ij}_{(2)} + \dots
}{eq:ap3}
and its determinant
\eq{
\sqrt{-\ga}=e^{2\hat\rho/\ell}\,\sqrt{-\ga^{(0)}} + \dots
}{eq:ap3a}
In all expressions above and below we display the leading and next-to-leading order terms (if they are non-vanishing) in powers of $e^{2\hat{\rho}/\ell}$.
The extrinsic curvature
\eq{
K_{ij} = \frac12 \,\partial_{\hat\rho} \ga_{ij} =e^{2\hat\rho/\ell}\, \frac1\ell \,\ga^{(0)}_{ij}+\frac{1}{2\ell}\,\ga_{ij}^{(1)} + \dots
}{eq:ap4}
and its inverse
\eq{
K^{ij} = e^{-2\hat\rho/\ell}\,\frac1\ell\,\ga^{ij}_{(0)}-e^{-4\hat\rho/\ell}\,\frac{2\hat\rho}{\ell^2}\,\ga^{ij}_{(1)}-e^{-4\hat\rho/\ell}\,\frac{2}{\ell}\,\ga^{ij}_{(2)}+e^{-4\hat\rho/\ell}\,\frac{1}{2\ell}\,\ga^{ij}_{(1)}+\dots
}{eq:ap4a}
in our case have a very simple trace
\eq{
K = \frac{2}{\ell} + \dots
}{eq:ap5}
because of the tracelessness gauge conditions [cf.~\eqref{eq:cg35}]
\eq{
\ga^{ij}_{(0)}\ga^{(1)}_{ij}=\ga^{ij}_{(0)}\ga^{(2)}_{ij}=0\,.
}{eq:ap6}
The Gauss-Codazzi equations
\eq{
R^i{}_{\hat\rho j\hat\rho} = -\partial_{\hat\rho} K^i{}_j - K^i{}_k K^k{}_j 
}{eq:ap7}
yield
\eq{
R^{k\hat\rho}{}_{j\hat\rho} = -\frac{1}{\ell^2}\,\de^k_j +e^{-2\hat\rho/\ell}\,\frac{1}{\ell^2}\,\ga^{kl}_{(1)}\ga^{(0)}_{lj}+\dots 
}{eq:ap8}

Analogous formulas are valid for the variations of these quantities. We use them to derive 
\eq{
\eps^{ij}R^{k\hat\rho}{}_{j\hat\rho}\,\de\ga_{ik}=\frac{1}{\ell^2}\,\eps^{ij}\,\ga^{kl}_{(1)}\ga_{lj}^{(0)}\,\de\ga_{ik}^{(0)}+\dots
}{eq:ap9}
and
\eq{
\eps^{ij} \,K_i{}^k\,\de K_{kj}=-\frac1\ell\,\eps^{ij}\,\Big(\frac{\hat\rho}{\ell^2}\,\ga^{lk}_{(1)}\ga_{li}^{(0)}+\frac1\ell\,\ga_{(2)}^{lk}\ga_{li}^{(0)}-\frac{1}{2\ell}\,\ga_{(0)}^{lk}\ga_{li}^{(1)}\Big)\,\de\ga_{kj}^{(0)}+\dots
}{eq:ap10}
In these expressions
\eq{
\eps^{ij} = \frac{\epsilon^{ij}}{\sqrt{-\ga^{(0)}}}
}{eq:apnew}
denotes the $\eps$-tensor with respect to the conformal boundary metric $\ga^{(0)}$. For the Einstein-Hilbert part of the action we need the quantity
\eq{
\sqrt{-\ga}\,\Big(K^{ij}-\big(K-\frac1\ell\big)\ga^{ij}\Big)\,\de\ga_{ij} = -\sqrt{-\ga^{(0)}}\Big(\frac{\hat\rho}{\ell^2}\,\ga^{ij}_{(1)}+\frac1\ell\,\ga^{ij}_{(2)}-\frac{1}{2\ell}\,\ga^{ij}_{(1)} \Big)\,\de\ga_{ij}^{(0)}+\dots 
}{eq:ap11}
The explicit form of the expression
$\Ga^k{}_{li}\,\de\Ga^l{}_{kj} \sim \ga^{(0)}\,\de\ga^{(0)}$
is not needed in the present work since it vanishes for flat $\ga^{(0)}$.
Dropping this term in the first variation of the on-shell action \eqref{eq:cg9} and using
\eq{
\hat\rho=\ell\rho 
}{eq:ap12}
establishes
\begin{align}
\de I_{\rm CG}\big|_{\rm EOM} = \lim_{\rho\to\infty}\int_{\dM}\!\!\!\!d^2x\sqrt{-\ga^{(0)}}\,\de\ga_{ij}^{(0)}\, \Big[\frac{\rho}{\ell}\big(\ga^{ij}_{(1)}-\ga^{il}_{(1)}\ga_{lk}^{(0)}\eps^{kj}\big)  \nonumber \\
 -\frac{1}{2\ell}\,\big(\ga^{ij}_{(1)}-3\ga^{il}_{(1)}\ga_{lk}^{(0)}\eps^{kj}\big) +\frac1\ell\,\big(\ga_{(2)}^{ij}-\ga^{il}_{(2)}\ga_{lk}^{(0)}\eps^{kj}\big)
\Big].
\label{eq:ap13}
\end{align}
The terms in the first line of \eqref{eq:ap13} diverge linearly with $\rho$, while the terms in the second line are finite. We see explicitly from \eqref{eq:ap13} that no $\de\ga^{(1)}$ or $\de\ga^{(2)}$ dependence remains for large $\rho$. Thus, the variational principle is well-defined.

\end{appendix}


\providecommand{\href}[2]{#2}\begingroup\raggedright\endgroup

\end{document}